\documentclass[journal]{IEEEtran}

\ifCLASSINFOpdf
\else
   \usepackage[dvips]{graphicx}
\fi
\usepackage{url}

\usepackage{graphicx}
\usepackage{cite}
\usepackage{booktabs}
\usepackage{hyperref}
\usepackage[dvipsnames, svgnames, x11names]{xcolor}
\usepackage{amsmath}
\usepackage{amssymb}
\usepackage{array}

\definecolor{mblu}{RGB}{31,119,180}
\definecolor{mred}{RGB}{255,127,13}
\definecolor{mgrn}{RGB}{44,160,44}
\definecolor{dred}{RGB}{250, 127, 111}
\definecolor{dgrn}{RGB}{90, 175, 52}
\definecolor{dyel}{RGB}{255, 190, 122}

\begin{document}

\title{Synthetic Speech Detection Based on Temporal Consistency and Distribution of Speaker Features}

\author{Yuxiang Zhang, Zhuo Li, Jingze Lu, Wenchao Wang, Pengyuan Zhang
\thanks{Corresponding author: Wenchao Wang and Pengyuan Zhang
	
	The authors are with Key Laboratory of Speech Acoustics and Content Understanding, Institute of Acoustics, Chinese Academy of Sciences, Beijing 100190, China and also with University of Chinese Academy of Sciences, Beijing 100049, China (e-mail: \{zhangyuxiang, lujingze, wangwenchao, zhangpengyuan\}@hccl.ioa.ac.cn and li\_zhuo@foxmail.com).}}

\markboth{Journal of \LaTeX\ Class Files, Vol. 14, No. 8, August 2015}
{Shell \MakeLowercase{\textit{et al.}}: Bare Demo of IEEEtran.cls for IEEE Journals}
\maketitle

\begin{abstract}
Current synthetic speech detection (SSD) methods perform well on certain datasets but still face issues of robustness and interpretability. A possible reason is that these methods do not analyze the deficiencies of synthetic speech. In this paper, the flaws of the speaker features inherent in the text-to-speech (TTS) process are analyzed. Differences in the temporal consistency of intra-utterance speaker features arise due to the lack of fine-grained control over speaker features in TTS. Since the speaker representations in TTS are based on speaker embeddings extracted by encoders, the distribution of inter-utterance speaker features differs between synthetic and bonafide speech. Based on these analyzes, an SSD method based on temporal consistency and distribution of speaker features is proposed. On one hand, modeling the temporal consistency of intra-utterance speaker features can aid speech anti-spoofing. On the other hand, distribution differences in inter-utterance speaker features can be utilized for SSD. The proposed method offers low computational complexity and performs well in both cross-dataset and silence trimming scenarios.
\end{abstract}


\IEEEpeerreviewmaketitle

\section{Introduction}

\IEEEPARstart{T}{he} relationship between synthetic speech detection (SSD) and automatic speaker verification (ASV) has been long-standing. ASV systems have made significant progress with the development of deep learning. But they remain vulnerable to spoofing attacks~\cite{shchemelinin2013examining, kinnunen2012vulnerability, wu2015spoofing}. To address this vulnerability, the biennial ASVspoof challenges~\cite{wu2015asvspoof, Kinnunen2017, todisco2019asvspoof, yamagishi21_asvspoof} were introduced to promote the development of countermeasures (CMs) aim at protecting ASV systems from spoofing attacks. The Spoofing Aware Speaker Verification (SASV) challenge~\cite{jung22c_interspeech} goes further by considering integrated systems, where both CM and ASV systems are optimized to enhance reliability. Several approaches have been explored to incorporate speaker information into SSD, such as joint optimization of ASV systems and anti-spoofing CMs through multi-task~\cite{li2019multi, zhao2022multi} and joint-decision~\cite{todisco2018integrated, kanervisto2021optimizing, choi22b_interspeech, wang22ea_interspeech}. Furthermore, efforts have been made to implement \cite{alenin22_interspeech, zhang22s_interspeech, 9975428} or improve~\cite{ma2023boost} SSD based on ASV systems. But there remains a dearth of research on the relationship between SSD and ASV.

Most existing SSD systems, with the exception of end-to-end detectors such as RawNet2~\cite{9414234, tak21_asvspoof} and \mbox{AASIST}~\cite{9747766}, can be divided into two main components: feature extraction and classifier. Commonly used features include short-time Fourier transform (STFT) spectrogram, const Q cepstral coefficients (CQCC)~\cite{todisco2017constant}, linear frequency cepstral coefficients (LFCC)~\cite{sahidullah15_interspeech}, and others. Neural network classifiers typically rely on convolutional neural networks (CNN) such as light convolutional neural network (LCNN)~\cite{lavrentyeva2019stc, wang21fa_interspeech} and residual network (ResNet)~\cite{lai2019assert, li2021replay}. Alhough current CMs perform well on certain datasets, they lack interpretability and struggle to generalize well in open-set scenarios or when silence is absent~\cite{zhang21da_interspeech, muller21_asvspoof, 9746204, muller22_interspeech}. This limitation is partly due to the inadequate analysis of inherent flaws in the text-to-speech (TTS) process. And addressing the realistic representation of speakers is a common challenge in speech synthesis.

In TTS algorithms, speaker representations are typically fixed as one-hot embeddings or d-vectors \cite{wang2020asvspoof} obtained from specific speaker encoders. The loss of information during encoding and residual source speaker information in the voice conversion (VC) algorithms may result in differences in the distribution of speaker features when compared to bonafide speech \cite{ma2023boost, Gao2023}. Additionally, synthetic speech exhibits higher temporal consistency (TC) in intra-utterance speaker embeddings due to the invariant speaker representation. In contrast, bonafide speech demonstrates lower TC in intra-utterance speaker features due to variations in speaker state, such as vocal tract shape and speaking style. 
Therefore, the TC of intra-utterance speaker embeddings can also serve as a cue for SSD. While there are approaches based on temporal identity consistency in face deepfake detection \cite{liu2023ti2net}, no similar method has been developed for SSD. 

In this paper, the differences in the distribution of inter-utterance speaker features and the TC of the intra-utterance speaker features between synthetic and bonafide speech are first illustrated. 
The complementarity between the distribution and TC of speaker features in SSD is also demonstrated. Finally, based on the analysis of flaws in speaker representation in synthetic speech, an SSD method that leverages TC and the distribution of speaker features is proposed. 
The experimental results show that our algorithm achieves comparable performance with a relatively small number of parameters and low computational complexity. It also improves generalization across datasets and robustness to silence trimming scenario.

\section{Method}\label{sec2}
This section analyzes the differences and complementarity in intra-utterance TC and inter-utterance feature distribution between synthetic and bonafide speech. The model architecture for SSD is also proposed.

\subsection{Temporal Consistency of Intra-utterance Speaker Features}
\begin{figure}[tbp]
	\centering
	\includegraphics[width=0.97\columnwidth]{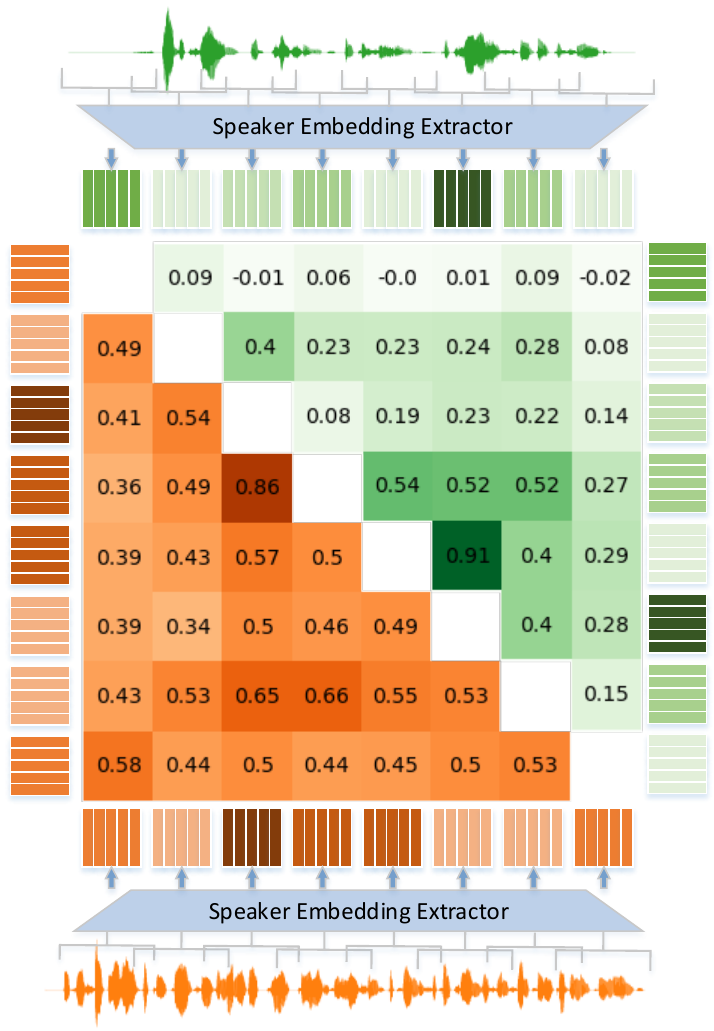}
	\caption{Temporal consistency of intra-utterance speaker features for synthetic speech and bonafide speech. The values in the lower triangle represent the similarity of the intra-utterance speaker embedding for \textcolor{mred}{spoof} speech LA\_T\_1477244, while the values in the upper triangle represent that for \textcolor{mgrn}{bonafide} speech LA\_T\_4155454.}
	\label{tc}
\end{figure}

There are differences in the TC of speaker features between synthetic and bonafide speech due to the difficulty of controlling speaker representations at a fine-grained level during TTS. ASVspoof 2019 LA dataset encompasses three types of speaker representation: d-vector, one-hot embedding, and variational auto-encoder (VAE) \cite{wang2020asvspoof}. They also cover a wide range of speaker representations in expressive TTS \cite{Tan2021ASO}. These embeddings are obtained in advance with speaker embedding encoders, encoding each speaker individually without considering variations in emotion, scene or style, etc. However, the intra-utterance short-time speaker embeddings of bonafide speech are highly variable because the speaker state varies with time, environment, and other factors. As a result, the TC of intra-utterance speaker embeddings is lower in bonafide speech compared to synthetic speech. Figure~\ref{tc} illustrates the difference in TC of intra-utterance speaker embeddings between bonafide and synthetic speech. The speech examples are derived from speaker LA\_0079 in ASVspoof 2019 LA training partition. Speaker embeddings of eight randomly selected 0.5s segments from each speech are separately extracted and arranged chronologically to calculate the cosine similarities between the intra-utterance speaker embeddings.

The intra-utterance speaker embeddings of synthetic speech have high similarities and a narrow range of variation In the lower triangle of the similarity matrix. Conversely, the upper triangle representing bonafide speech shows lower similarities and a wider range of variation. This indicates that there are differences in the TC of intra-utterance speaker embedding between synthetic and bonafide speech.


\subsection{Distribution of Inter-utterance Speaker Embedding}\label{sec2.2}
\begin{figure}[htbp]
	\centering
	\includegraphics[width=0.9\columnwidth]{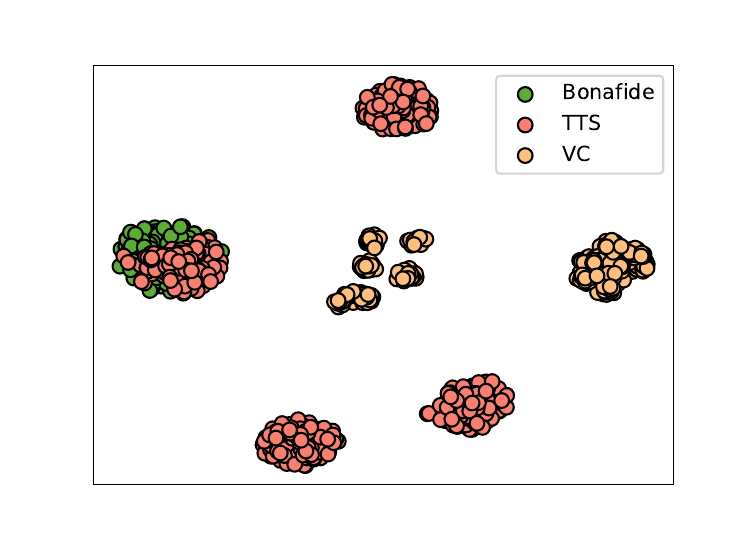}
	\caption{Visualization of differences in the distribution of inter-utterance speaker embeddings from speaker LA\_0079. \textcolor{dgrn}{Green}: Bonafide speech. \textcolor{dred}{Red}: TTS attacks. \textcolor{dyel}{Yellow}: VC attacks.}
	\label{dis}
\end{figure}

Current target person-oriented TTS and voice conversion (VC) algorithms are capable of generating high-quality synthetic speech~\cite{Tan2021ASO}. However, speaker embeddings differ between spoof and bonafide speech due to the loss of speaker encoding in spoofing algorithms. In particular, VC algorithms preserve information about the source speaker \cite{10096733, Cai2022InvertibleVC}. Consequently, the speaker embeddings of spoof speech, generated by VC algorithms, differ significantly from the embeddings of bonafide speech. Figure~\ref{dis} illustrates the discrepancy in the distribution of speaker embeddings between the two categories--spoof speech and bonafide speech. The t-distributed Stochastic Neighbor Embedding (t-SNE) \cite{van2008visualizing} is applied to highlight the differences in the distribution. Speaker embeddings are extracted from 1,243 utterances of speaker LA\_0079 in ASVspoof 2019 LA training set. The training set consists 4 TTS and 2 VC algorithms. 

The visualization of the dimension-reduced embedding clearly highlights significant differences in the distribution of speaker embeddings between bonafide and spoof speech, particularly for the spoof speech generated by the VC algorithms. While the distribution of some embeddings for the spoof speech generated by the TTS algorithms resambles that of bonafide speech. But there are differences in the distribution of the majority of embeddings.

In summary, the speaker features of bonafide speech differ in both TC and feature distribution compared to synthetic speech. Specifically, the TC of speaker features within utterances is lower for TTS-generated speech, whereas VC-generated speech has greater differences in inter-utterance speaker feature distribution.

\subsection{Model Architecture}
Based on the analysis of the differences in TC and distribution of speaker features between synthetic and bonafide speech, an SSD method composed of two parts is proposed.

CM1 performs anti-spoofing through the differences in temporal TC of speaker features between bonafide and synthetic speech. The TC of intra-utterance speaker features is modeled using the difference $\boldsymbol{D}_i$ of the speaker features $\boldsymbol{S}_i$, whichis fed into two layers of Gated Recurrent Unit (GRU). To capture fine-grained TC, the speaker features are obtained from the multi-layer feature aggregation (MFA) of the pre-trained emphasized channel attention, propagation, and aggregation time delay neural network (ECAPA-TDNN) ASV system instead of using embeddings. 
To eliminate the influence of speaker identity and channel within the speaker features and emphasize TC differences, the differences $\boldsymbol{D}_i$ between adjacent frames in $\boldsymbol{S}_i$ are calculated in advance:
\begin{equation}
	\begin{aligned}
		\boldsymbol{D}_i&=\{\boldsymbol{d}_{i,1}, \boldsymbol{d}_{i,2}, \dots, \boldsymbol{d}_{i,t-1}\}\\
		&=\{\boldsymbol{s}_{i,2}-\boldsymbol{s}_{i,1}, \boldsymbol{s}_{i,3}-\boldsymbol{s}_{i,2}, \dots, \boldsymbol{s}_{i,t}-\boldsymbol{s}_{i,t-1}\}
	\end{aligned} \nonumber
\end{equation}
Finally, the last output of the GRU is fed into two FC layers for binary classification.

The proposed CM2 focuses on the differences in the distribution of inter-utterance speaker embeddings is similar to our previous work\cite{zhang22s_interspeech}. The architecture of ECAPA-TDNN remains unchanged, with the feature extractor preceding the MFA remaining fixed and the subsequent layers are updated.

\begin{figure}[htbp]
	\centering
	\includegraphics[width=1.02\columnwidth]{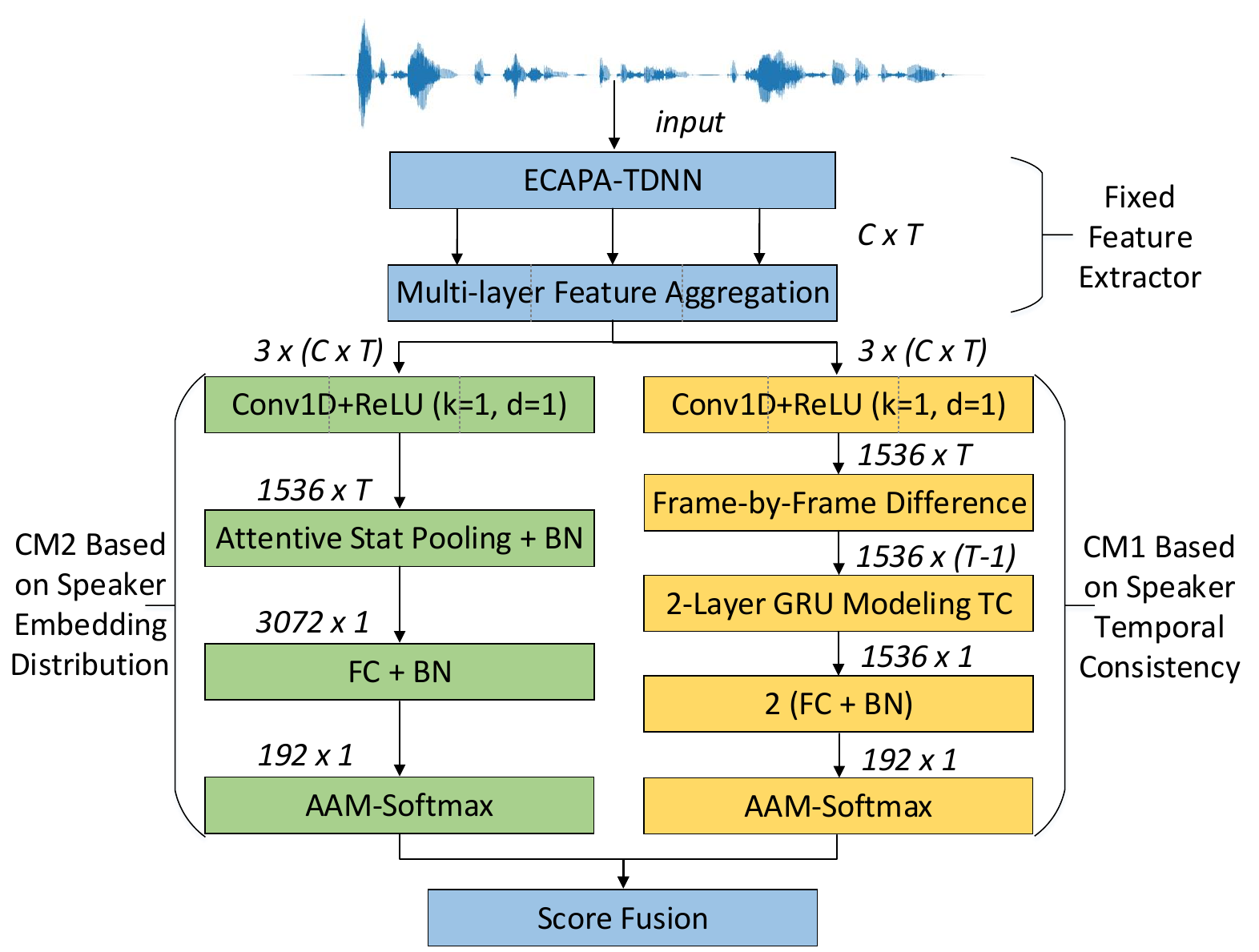}
	\caption{Model architecture of proposed SSD method.}
	\label{model}
\end{figure}

Figure \ref{model} shows the model architecture of the proposed SSD method based on speaker features. Both CMs are optimized with the Additive Angular Margin Softmax (AAM-Softmax) loss function \cite{deng2019arcface}. Leveraging the complementarity of the features described in Section \ref{sec2.2}, the scores from the two anti-spoofing systems are fused with equal weights of $0.5$.

In addition to exploit the differences in speaker features between synthetic and bonafide speech, there are other advantages to extracting features from pre-trained ASV systems. Firstly, the pre-trained ASV systems are trained for the ASV task, so the features extracted are able to mitigate the influence of information such as speech content. Additionally, pre-trained ASV systems are trained on extensive bonafide speech datasets, resulting in improved anti-spoofing robustness. Moreover, the proposed methods involve fewer trainable parameters than common pre-trained models such as wav2vec \cite{baevski2020wav2vec}. The proposed SSD method based on pre-trained ASV systems can also be more compactly integrated with ASV systems.

\section{Experiments}
\subsection{Datasets and Evaluation Metrics}
ASVspoof 2019 LA dataset \cite{wang2020asvspoof} has been widely utilized in speech anti-spoofing since its release. Thus all experiments presented here are couducted on this dataset (19LA). Only the training partition is used for model training. 
To validate the generalization of the proposed methods, we evaluate the cross-channel and cross-dataset performance on ASVspoof 2021 LA (21LA) \cite{yamagishi21_asvspoof} and IntheWild dataset \cite{muller22_interspeech}. Additionally, we assess the robustness of the proposed CMs to VAD by evaluating them on 21LA hidden track (21hid) and 19LA evaluation partition with silence trimmed (19trim). The silence segments are trimmed via librosa.effects.trim \footnote{\url{http://librosa.org/doc/latest/generated/librosa.effects.trim.html}} with the \textit{top\_db} of 40. The equal error rate (EER) is used as the evaluation metric.

\subsection{Training Details}
The pre-trained ECAPA-TDNN model is provided by the SASV baseline\footnote{\url{https://github.com/TaoRuijie/ECAPA-TDNN}}.
The input speech duration ranges from 2 to 4 seconds. The input features are 80-dimensional FBank with Hamming window of 400 frame length and 512 Fast Fourier Transform (FFT) points. The SpecAugment is used for data augmentation. The number of the GRU hidden state is 1536, while the two linear classification layers have output sizes of 512 and 192. The parameters of the AAM-Softmax loss function are $m=0.4, s=30$. The Adam optimizer 
is applied to update the model. The learning rate increases linearly to $3e^{-4}$ for the first 1000 warm-up steps and then decreases proportionally to the inverse square root of the step number. Training is conducted over 30 epochs and a batchsize of 256. 

\subsection{Results}
\subsubsection{Evaluation on ASVspoof datasets}
Table \ref{tab2} presents the EERs of the TTS and VC algorithms in ASVspoof 2019 LA datasets, validating the complementarity between the proposed two CMs. Although both systems exhibit better performance in detecting TTS algorithms, the results of the evaluation set demonstrate the complementarity of the two systems for detecting different kinds of spoofing algorithms. The TC-based CM1 shows superior performance against VC algorithms, while the CM2 based on the differences in the distribution of speaker features obtains a lower EER in detecting the TTS algorithms. 
Benefiting from the complementarity, the score fusion yeilds a relative performance improvement of 37\% on the evaluation partition.


\begin{table}[htbp]
	\caption{EER\% of the proposed CMs for different types of spoofing algorithms in ASVspoof 2019 LA datasets}
	\centering
	\setlength{\tabcolsep}{3.0pt}
	\begin{tabular}{c | c c c | c c c}
		\bottomrule
		Systems & Dev TTS & Dev VC & Dev all & Eval TTS & Eval VC & Eval all \\
		\hline
		CM1 & 0.39 & 0.66 & 0.51 & 1.51 & \textbf{2.41} & 1.79 \\
		CM2 & 0.28 & 0.39 & 0.32 & \textbf{1.25} & 3.07 & 1.81 \\
		Fusion & 0.12 & 0.20 & \textbf{0.16} & 0.78 & 1.55 & \textbf{1.14} \\
		\toprule
	\end{tabular}
	\label{tab2}
\end{table}

To demonstrate the effectiveness and cross-channel robustness of the proposed SSD methods, the CMs are evaluated on the 19LA and 21LA evaluation subsets. The experimental results are compared with some of the latest state-of-the-art (SOTA) systems and are shown in Table \ref{tab3}. 
The performance of the fusion system is comparable to current SOTA systems on both evaluation datasets. However, the performance of proposed CMs is slightly inferior to that of DFSincNet \cite{10057965}. 
The CM1 performs as well as AASIST and wav2vec based SSD systems on 21LA. The CM2 without GRU has lower computational complexity than that of AASIST and a lower EER on 21LA. Compared to the wav2vec 2.0 based CM without data augmentation, which is also based on a pre-trained model, our pre-trained ASV system based CMs perform better. The proposed method also has a much smaller number of trainable parameters and floating point operations (FLOPs). This is probably because only some of the layers are trained in our method, whereas the wav2vec-based method requires fine-tuning of all layers.

\begin{table}[htbp]
	\caption{EER\% comparison with existing SOTA systems on ASVspoof 2019 LA and 2021 LA evaluation datasets}
	\centering
	\begin{tabular}{c c c c c}
		\toprule
		System & Trainable Params & FLOPs & 19LA & 21LA \\
		\midrule
		DFSincNet\cite{10057965} & 0.43 M & \textbf{16.47 G} & \textbf{0.52} & \textbf{3.38}\\
		AASIST\cite{9747766} & \textbf{0.30 M} & 19.41 G & 0.83 & 7.65 \\
		Wav2vec\cite{10094779}& 317.84 M & 150.39 G & 2.98 & 7.53 \\
		\midrule
		CM1 & 32.37 M & 24.67 G & 1.79 & 7.65\\
		CM2 & 6.57 M & \textbf{8.51 G} & 1.81 & 6.38\\
		Fusion & 38.94 M & 28.49 G & \textbf{1.14} & \textbf{5.38}\\
		\bottomrule
	\end{tabular}
	\label{tab3}
\end{table}

\subsubsection{Demonstrating the Robustness}
To validate the cross-dataset robustness of the proposed SSD methods, the CMs are evaluated on the IntheWild dataset.
The results are reported in Table \ref{tab4}. The proposed anti-spoofing CM1 based on TC of speaker features gains better cross-dataset performance than CMs reported in \cite{muller22_interspeech}. This can possibly be attributed to the TC differences between bonafide and spoof speech that are also present in this dataset. Furthermore, the fusion system achieves comparable robustness to the wav2vec based system, while requiring fewer trainable parameters and FLOPs.

\begin{table}[htbp]
	\caption{Cross-dataset evaluation on the In the Wild dataset}
	\centering
	\begin{tabular}{c c c c}
		\toprule
		System & Trainable Params & FLOPs & EER\% \\
		\midrule
		RawGAT-ST \cite{muller22_interspeech} & 0.43 M & 37.61 G & 37.15 \\
		RawNet2 \cite{muller22_interspeech} & 25.43 M & 3.73 G & 33.94 \\
		Wav2vec \cite{10094779} & 317.84 M & 150.39 G & \textbf{26.65} \\
		\midrule
		CM1 & 32.37 M & 24.67 G & 29.66\\
		CM2 & 6.57 M & 8.51 G & 31.10\\
		Fusion & 38.94 M & 28.49 G & \textbf{27.30}\\
		\bottomrule
	\end{tabular}
	\label{tab4}
\end{table}

The results on the 19trim and 21hid datasets demonstrate the robustness of the proposed methods to VAD, as shown in Table \ref{tab5}. CM2 obtains the lowest EER on both evaluation datasets when the non-speech segments at the beginning and end are trimmed. This suggests that the proposed anti-spoofing CM2 based on the differences in the distribution of speaker embeddings exhibits better robustness in detecting spoof speech with silence removed. One possible reason for the good robustness is that the differences in the distribution of speaker features, which are extracted from a pre-trained ASV system, are less affected by silence.

\begin{table}[htbp]
	\caption{EER\% comparison of evaluation of datasets after VAD}
	\centering
	\begin{tabular}{c c c c c}
		\toprule
		System & Trainable Params & FLOPs & 19trim & 21hid \\
		\midrule
		Wav2vec\cite{10094779} & 317.84 M & 150.39 G & 15.56 & 28.80 \\
		\midrule
		CM1 & 32.37 M & 24.67 G & 23.82 & 32.19\\
		CM2 & 6.57 M & 8.51 G & \textbf{14.56} & \textbf{27.60} \\
		Fusion & 38.94 M & 28.49 & 17.13 & 28.11\\
		\bottomrule
	\end{tabular}
	\label{tab5}
\end{table}

\section{Conclusioin}
This paper focuses on analyzing and utilizing the differences in inter- and intra-utterance speaker features between synthetic and bonafide speech for SSD. Firstly, differences in the TC of the intra-utterance speaker features between bonafide and spoof speech are demonstrated. The differences arise from the limited control over speaker features during TTS process. Additionally, the differences in the distribution of inter-utterance speaker features between bonafide and synthetic speech are shown. The differences in speaker feature distribution may be caused by residual source speaker information in VC and loss in speaker encoding. Based on these analyzes, an SSD method utilize both inter- and intra-utterance speaker features are proposed. The speaker features are extracted after the MFA through a pre-trained ECAPA-TDNN ASV system. The CM1 models the TC of the differenced speaker features by GRU and then performs speech anti-spoofing. The inter-utterance speaker feature based CM2 directly trains the subsequent layers after MFA for SSD. To further improve performance, score fusion is employed, as the two CMs are designed to detect different types of spoofing algorithms.
The proposed method has a small number of parameters, low computational complexity, and good performance. It is also robust to cross-dataset and VAD scenarios.

\bibliographystyle{IEEEtran}
\bibliography{mybib}

\begin{thebibliography}{10}
\providecommand{\url}[1]{#1}
\csname url@samestyle\endcsname
\providecommand{\newblock}{\relax}
\providecommand{\bibinfo}[2]{#2}
\providecommand{\BIBentrySTDinterwordspacing}{\spaceskip=0pt\relax}
\providecommand{\BIBentryALTinterwordstretchfactor}{4}
\providecommand{\BIBentryALTinterwordspacing}{\spaceskip=\fontdimen2\font plus
\BIBentryALTinterwordstretchfactor\fontdimen3\font minus
  \fontdimen4\font\relax}
\providecommand{\BIBforeignlanguage}[2]{{%
\expandafter\ifx\csname l@#1\endcsname\relax
\typeout{** WARNING: IEEEtran.bst: No hyphenation pattern has been}%
\typeout{** loaded for the language `#1'. Using the pattern for}%
\typeout{** the default language instead.}%
\else
\language=\csname l@#1\endcsname
\fi
#2}}
\providecommand{\BIBdecl}{\relax}
\BIBdecl

\bibitem{shchemelinin2013examining}
V.~Shchemelinin and K.~Simonchik, ``Examining vulnerability of voice
  verification systems to spoofing attacks by means of a tts system,'' in
  \emph{International Conference on Speech and Computer}.\hskip 1em plus 0.5em
  minus 0.4em\relax Springer, 2013, pp. 132--137.

\bibitem{kinnunen2012vulnerability}
T.~Kinnunen, Z.-Z. Wu, K.~A. Lee, F.~Sedlak, E.~S. Chng, and H.~Li,
  ``Vulnerability of speaker verification systems against voice conversion
  spoofing attacks: The case of telephone speech,'' in \emph{Proc. ICASSP
  2012}.\hskip 1em plus 0.5em minus 0.4em\relax IEEE, 2012, pp. 4401--4404.

\bibitem{wu2015spoofing}
Z.~Wu, N.~Evans, T.~Kinnunen, J.~Yamagishi, F.~Alegre, and H.~Li, ``Spoofing
  and countermeasures for speaker verification: A survey,'' \emph{speech
  communication}, vol.~66, pp. 130--153, 2015.

\bibitem{wu2015asvspoof}
Z.~Wu, T.~Kinnunen, N.~Evans, J.~Yamagishi, C.~Hanil{\c{c}}i, M.~Sahidullah,
  and A.~Sizov, ``Asvspoof 2015: the first automatic speaker verification
  spoofing and countermeasures challenge,'' in \emph{Proc. Interspeech 2015},
  2015, pp. 2037--2041.

\bibitem{Kinnunen2017}
T.~Kinnunen, M.~Sahidullah, H.~Delgado, M.~Todisco, N.~Evans, J.~Yamagishi, and
  K.~A. Lee, ``The asvspoof 2017 challenge: Assessing the limits of replay
  spoofing attack detection,'' in \emph{Proc. Interspeech 2017}, 2017, pp.
  2--6.

\bibitem{todisco2019asvspoof}
M.~Todisco, X.~Wang, V.~Vestman \emph{et~al.}, ``Asvspoof 2019: Future horizons
  in spoofed and fake audio detection,'' \emph{Proc. Interspeech 2019}, pp.
  1008--1012, 2019.

\bibitem{yamagishi21_asvspoof}
J.~Yamagishi, X.~Wang, M.~Todisco, M.~Sahidullah, J.~Patino, A.~Nautsch,
  X.~Liu, K.~A. Lee, T.~Kinnunen, N.~Evans, and H.~Delgado, ``{ASVspoof 2021:
  accelerating progress in spoofed and deepfake speech detection},'' in
  \emph{Proc. 2021 Edition of the Automatic Speaker Verification and Spoofing
  Countermeasures Challenge}, 2021, pp. 47--54.

\bibitem{jung22c_interspeech}
J.~weon Jung, H.~Tak, H.~jin Shim, H.-S. Heo, B.-J. Lee, S.-W. Chung, H.-J. Yu,
  N.~Evans, and T.~Kinnunen, ``Sasv 2022: The first spoofing-aware speaker
  verification challenge,'' in \emph{Proc. Interspeech 2022}, 2022, pp.
  2893--2897.

\bibitem{li2019multi}
J.~Li, M.~Sun, and X.~Zhang, ``Multi-task learning of deep neural networks for
  joint automatic speaker verification and spoofing detection,'' in \emph{2019
  Asia-Pacific Signal and Information Processing Association Annual Summit and
  Conference (APSIPA ASC)}.\hskip 1em plus 0.5em minus 0.4em\relax IEEE, 2019,
  pp. 1517--1522.

\bibitem{zhao2022multi}
Y.~Zhao, R.~Togneri, and V.~Sreeram, ``Multi-task learning-based
  spoofing-robust automatic speaker verification system,'' \emph{Circuits,
  Systems, and Signal Processing}, pp. 1--22, 2022.

\bibitem{todisco2018integrated}
M.~Todisco, H.~Delgado, K.~A. Lee, M.~Sahidullah, N.~Evans, T.~Kinnunen, and
  J.~Yamagishi, ``Integrated presentation attack detection and automatic
  speaker verification: Common features and gaussian back-end fusion,'' in
  \emph{Interspeech 2018}.\hskip 1em plus 0.5em minus 0.4em\relax International
  Speech Communication Association, 2018, pp. 77--81.

\bibitem{kanervisto2021optimizing}
A.~Kanervisto, V.~Hautam{\"a}ki, T.~Kinnunen, and J.~Yamagishi, ``Optimizing
  tandem speaker verification and anti-spoofing systems,'' \emph{IEEE/ACM
  Transactions on Audio, Speech, and Language Processing}, vol.~30, pp.
  477--488, 2021.

\bibitem{choi22b_interspeech}
J.-H. Choi, J.-Y. Yang, Y.-R. Jeoung, and J.-H. Chang, ``{HYU Submission for
  the SASV Challenge 2022: Reforming Speaker Embeddings with Spoofing-Aware
  Conditioning},'' in \emph{Proc. Interspeech 2022}, 2022, pp. 2873--2877.

\bibitem{wang22ea_interspeech}
X.~Wang, X.~Qin, Y.~Wang, Y.~Xu, and M.~Li, ``{The DKU-OPPO System for the 2022
  Spoofing-Aware Speaker Verification Challenge},'' in \emph{Proc. Interspeech
  2022}, 2022, pp. 4396--4400.

\bibitem{alenin22_interspeech}
A.~Alenin, N.~Torgashov, A.~Okhotnikov, R.~Makarov, and I.~Yakovlev, ``{A
  Subnetwork Approach for Spoofing Aware Speaker Verification},'' in
  \emph{Proc. Interspeech 2022}, 2022, pp. 2888--2892.

\bibitem{zhang22s_interspeech}
Y.~Zhang, Z.~Li, W.~Wang, and P.~Zhang, ``{SASV Based on Pre-trained ASV System
  and Integrated Scoring Module},'' in \emph{Proc. Interspeech 2022}, 2022, pp.
  4376--4380.

\bibitem{9975428}
A.~Pianese, D.~Cozzolino, G.~Poggi, and L.~Verdoliva, ``Deepfake audio
  detection by speaker verification,'' in \emph{2022 IEEE International
  Workshop on Information Forensics and Security (WIFS)}, 2022, pp. 1--6.

\bibitem{ma2023boost}
X.~Ma, S.~Zhang, S.~Huang, J.~Gao, Y.~Hu, and L.~He, ``How to boost
  anti-spoofing with x-vectors,'' in \emph{2022 IEEE Spoken Language Technology
  Workshop (SLT)}.\hskip 1em plus 0.5em minus 0.4em\relax IEEE, 2023, pp.
  593--598.

\bibitem{9414234}
H.~Tak, J.~Patino, M.~Todisco \emph{et~al.}, ``End-to-end anti-spoofing with
  rawnet2,'' in \emph{Proc. ICASSP 2021}, 2021, pp. 6369--6373.

\bibitem{tak21_asvspoof}
H.~Tak, J.~weon Jung \emph{et~al.}, ``{End-to-end spectro-temporal graph
  attention networks for speaker verification anti-spoofing and speech deepfake
  detection},'' in \emph{Proc. 2021 Edition of the Automatic Speaker
  Verification and Spoofing Countermeasures Challenge}, 2021, pp. 1--8.

\bibitem{9747766}
J.-w. Jung, H.-S. Heo, H.~Tak \emph{et~al.}, ``Aasist: Audio anti-spoofing
  using integrated spectro-temporal graph attention networks,'' in \emph{Proc.
  ICASSP 2022}, 2022, pp. 6367--6371.

\bibitem{todisco2017constant}
M.~Todisco, H.~Delgado, and N.~Evans, ``Constant q cepstral coefficients: A
  spoofing countermeasure for automatic speaker verification,'' \emph{Computer
  Speech \& Language}, vol.~45, pp. 516--535, 2017.

\bibitem{sahidullah15_interspeech}
M.~Sahidullah, T.~Kinnunen, and C.~Hanilçi, ``{A comparison of features for
  synthetic speech detection},'' in \emph{Proc. Interspeech 2015}, 2015, pp.
  2087--2091.

\bibitem{lavrentyeva2019stc}
G.~Lavrentyeva, A.~Tseren, M.~Volkova, A.~Gorlanov, A.~Kozlov, and
  S.~Novoselov, ``Stc antispoofing systems for the asvspoof2019 challenge,'' in
  \emph{Proc. Interspeech 2019}, 2019, pp. 1033--1037.

\bibitem{wang21fa_interspeech}
X.~Wang and J.~Yamagishi, ``{A Comparative Study on Recent Neural Spoofing
  Countermeasures for Synthetic Speech Detection},'' in \emph{Proc. Interspeech
  2021}, 2021, pp. 4259--4263.

\bibitem{lai2019assert}
C.-I. Lai, N.~Chen, J.~Villalba, and N.~Dehak, ``Assert: Anti-spoofing with
  squeeze-excitation and residual networks,'' in \emph{Interspeech}, 2019, pp.
  1013--1017.

\bibitem{li2021replay}
X.~Li, N.~Li, C.~Weng \emph{et~al.}, ``Replay and synthetic speech detection
  with res2net architecture,'' in \emph{Proc. ICASSP 2021}.\hskip 1em plus
  0.5em minus 0.4em\relax IEEE, 2021, pp. 6354--6358.

\bibitem{zhang21da_interspeech}
Y.~Zhang, W.~Wang, and P.~Zhang, ``{The Effect of Silence and Dual-Band Fusion
  in Anti-Spoofing System},'' in \emph{Proc. Interspeech 2021}, 2021, pp.
  4279--4283.

\bibitem{muller21_asvspoof}
N.~Müller, F.~Dieckmann \emph{et~al.}, ``{Speech is Silver, Silence is Golden:
  What do ASVspoof-trained Models Really Learn?}'' in \emph{Proc. 2021 Edition
  of the Automatic Speaker Verification and Spoofing Countermeasures
  Challenge}, 2021, pp. 55--60.

\bibitem{9746204}
X.~Wang and J.~Yamagishi, ``Estimating the confidence of speech spoofing
  countermeasure,'' in \emph{Proc. ICASSP 2022}, 2022, pp. 6372--6376.

\bibitem{muller22_interspeech}
N.~Müller, P.~Czempin, F.~Diekmann, A.~Froghyar, and K.~Böttinger, ``{Does
  Audio Deepfake Detection Generalize?}'' in \emph{Proc. Interspeech 2022},
  2022, pp. 2783--2787.

\bibitem{wang2020asvspoof}
X.~Wang, J.~Yamagishi, M.~Todisco, H.~Delgado, A.~Nautsch, N.~Evans,
  M.~Sahidullah, V.~Vestman, T.~Kinnunen, K.~A. Lee \emph{et~al.}, ``Asvspoof
  2019: A large-scale public database of synthesized, converted and replayed
  speech,'' \emph{Computer Speech \& Language}, vol.~64, p. 101114, 2020.

\bibitem{Gao2023}
Y.~Gao, ``{Audio Deepfake Detection Based on Differences in Human and Machine
  Generated Speech},'' Ph.D. dissertation, Carnegie Mellon University, 5 2022.

\bibitem{liu2023ti2net}
B.~Liu, B.~Liu, M.~Ding, T.~Zhu, and X.~Yu, ``Ti2net: Temporal identity
  inconsistency network for deepfake detection,'' in \emph{Proceedings of the
  IEEE/CVF Winter Conference on Applications of Computer Vision}, 2023, pp.
  4691--4700.

\bibitem{Tan2021ASO}
X.~Tan, T.~Qin, F.~K. Soong, and T.-Y. Liu, ``A survey on neural speech
  synthesis,'' \emph{ArXiv}, vol. abs/2106.15561, 2021.

\bibitem{10096733}
D.~Cai, Z.~Cai, and M.~Li, ``Identifying source speakers for voice conversion
  based spoofing attacks on speaker verification systems,'' in \emph{Proc.
  ICASSP 2023}, 2023, pp. 1--5.

\bibitem{Cai2022InvertibleVC}
Z.~Cai and M.~Li, ``Invertible voice conversion,'' \emph{ArXiv}, vol.
  abs/2201.10687, 2022.

\bibitem{van2008visualizing}
L.~Van~der Maaten and G.~Hinton, ``Visualizing data using t-sne.''
  \emph{Journal of machine learning research}, vol.~9, no.~11, 2008.

\bibitem{deng2019arcface}
J.~Deng, J.~Guo, N.~Xue \emph{et~al.}, ``Arcface: Additive angular margin loss
  for deep face recognition,'' in \emph{Proceedings of the IEEE/CVF conference
  on computer vision and pattern recognition}, 2019, pp. 4690--4699.

\bibitem{baevski2020wav2vec}
A.~Baevski, Y.~Zhou, A.~Mohamed, and M.~Auli, ``wav2vec 2.0: A framework for
  self-supervised learning of speech representations,'' \emph{Advances in
  neural information processing systems}, vol.~33, pp. 12\,449--12\,460, 2020.

\bibitem{10057965}
B.~Huang, S.~Cui, J.~Huang, and X.~Kang, ``Discriminative frequency information
  learning for end-to-end speech anti-spoofing,'' \emph{IEEE Signal Processing
  Letters}, vol.~30, pp. 185--189, 2023.

\bibitem{10094779}
X.~Wang and J.~Yamagishi, ``Spoofed training data for speech spoofing
  countermeasure can be efficiently created using neural vocoders,'' in
  \emph{Proc. ICASSP 2023}, 2023, pp. 1--5.

\end{thebibliography}

\end{document}